\documentclass[aps,prb,twocolumn,superscriptaddress]{revtex4-1}

\usepackage{graphicx,epstopdf,siunitx,braket,xspace,amssymb,dsfont,bm,todonotes, natbib,amsmath,multirow}

\setcitestyle{super, sort} 
\begin{document}
	
\title{Tuning methods for semiconductor spin--qubits}

\author{Tim Botzem}
\affiliation{JARA-FIT Institute Quantum Information, Forschungszentrum J\"ulich GmbH and RWTH Aachen University, 52074 Aachen, Germany}
\author{Michael D. Shulman}
\affiliation{Department of Physics, Harvard University, Cambridge, MA, 02138, USA}
\author{Sandra Foletti}
\affiliation{Department of Physics, Harvard University, Cambridge, MA, 02138, USA}
\author{Shannon P. Harvey}
\affiliation{Department of Physics, Harvard University, Cambridge, MA, 02138, USA}
\author{Oliver E. Dial}
\affiliation{Department of Physics, Harvard University, Cambridge, MA, 02138, USA}
\author{Patrick Bethke}
\affiliation{JARA-FIT Institute Quantum Information, Forschungszentrum J\"ulich GmbH and RWTH Aachen University, 52074 Aachen, Germany}	
\author{Pascal Cerfontaine}
\affiliation{JARA-FIT Institute Quantum Information, Forschungszentrum J\"ulich GmbH and RWTH Aachen University, 52074 Aachen, Germany}	
\author{Robert P. G. McNeil}
\affiliation{JARA-FIT Institute Quantum Information, Forschungszentrum J\"ulich GmbH and RWTH Aachen University, 52074 Aachen, Germany}	
\author{Diana Mahalu}
\affiliation{Braun Center for Submicron Research, Department of Condensed Matter Physics,
	Weizmann Institute of Science, Rehovot 76100, Israel}
\author{Vladimir Umansky}
\affiliation{Braun Center for Submicron Research, Department of Condensed Matter Physics,
	Weizmann Institute of Science, Rehovot 76100, Israel}
\author{Arne Ludwig}
\affiliation{Lehrstuhl f\"ur Angewandte Festk\"orperphysik, Ruhr-Universit\"at Bochum, D-44780 Bochum, Germany}
\author{Andreas Wieck}
\affiliation{Lehrstuhl f\"ur Angewandte Festk\"orperphysik, Ruhr-Universit\"at Bochum, D-44780 Bochum, Germany}
\author{Dieter Schuh}
\affiliation{Institut f\"ur Experimentelle und Angewandte Physik, Universit\"at Regensburg, D-93040 Regensburg, Germany}
\author{Dominique Bougeard}
\affiliation{Institut f\"ur Experimentelle und Angewandte Physik, Universit\"at Regensburg, D-93040 Regensburg, Germany}
\author{Amir Yacoby}
\affiliation{Department of Physics, Harvard University, Cambridge, MA, 02138, USA}
\author{Hendrik Bluhm}
\affiliation{JARA-FIT Institute Quantum Information, Forschungszentrum J\"ulich GmbH and RWTH Aachen University, 52074 Aachen, Germany}
\date{\today}
	
\begin{abstract}
We present efficient methods to reliably characterize and tune gate-defined semiconductor spin qubits. Our methods are designed to target the tuning procedures of semiconductor double quantum dot in GaAs heterostructures, but can easily be adapted to other quantum-dot-like qubit systems. These tuning procedures include the characterization of the inter-dot tunnel coupling, the tunnel coupling to the surrounding leads and the identification of the various fast initialization points for the operation of the qubit.  Since semiconductor-based spin qubits are compatible with standard semiconductor process technology and hence promise good prospects of scalability, the challenge of efficiently tuning the dot's parameters will only grow in the near future, once the multi-qubit stage is reached. With the anticipation of being used as the basis for future automated tuning protocols, all measurements presented here are fast-to-execute and easy-to-analyze characterization methods. They result in quantitative measures of the relevant qubit parameters within a couple of seconds, and require almost no human interference. 
\end{abstract}
\maketitle
\section{Introduction}
\label{ch:tuning:intro}
The recent developments in semiconductor based spin qubits show their great potential as building blocks of a quantum computer and demonstrate their promise for scalable architectures.\cite{cerfontaine_feedback-tuned_2016-1, baart_single-spin_2016,kawakami_electrical_2014,muhonen_quantifying_2015,maune_coherent_2012,takakura_single_2014,delbecq_full_2014,veldhorst_addressable_2014,veldhorst_two-qubit_2015}  However, with increasing number of physical qubits, challenges like device architecture,\cite{macleod_hybrid_2015,frees_compressed_2014,mak_ultra-shallow_2013-1} long-range coupling,\cite{mcneil_-demand_2011,trifunovic_long-distance_2012,serina_long-range_2016} error correction,\cite{fowler_high-threshold_2009,fowler_surface_2012} decoherence due to charge noise,\cite{kuhlmann_charge_2013,dial_charge_2013} and scalable implementation\cite{johansson_qutip_2013,svore_layered_2006} of the control electronics\cite{jones_layered_2012,levy_implications_2011} will play an increasingly important role. One further obstacle, which has not received as much attention to date, is the tuning of the qubit devices. Especially in the case of gate-defined quantum dots, even tuning a single two--electron quibit is already a non-trivial task, as each quantum dot comprises at least three electrostatic gate--electrodes, each of which influences the number of electrons in the dot, the tunnel coupling to the adjacent lead, and the inter-dot tunnel coupling. The current practice of manually tuning the qubits is a time-consuming and non-scalable procedure.

In this work we present tuning and characterization methods used to realize two-electron spin qubits in GaAs, which have evolved over the course of the experiments presented in Refs.\,\onlinecite{cerfontaine_feedback-tuned_2016-1} and \onlinecite{bluhm_dephasing_2011,bluhm_enhancing_2010,shulman_demonstration_2012, dial_charge_2013, shulman_suppressing_2014, nichol_quenching_2015,botzem_quadrupolar_2016}. Complementary to Ref.\,\onlinecite{baart_computer-automated_2016}, which  shows a computer--automated scheme for the coarse-tuning of quantum dots into the single-electron regime, we focus here on the fine--tuning of the spin qubit once the single-electron regime is reached. This includes the adjustment and the characterization of the tunnel couplings to adjacent leads and between the dots, and the identification of the energy transitions relevant for the qubit functionality. We exploit high--bandwidth readout by radiofrequency (RF)-reflectometry\cite{schoelkopf_radio-frequency_1998, reilly_fast_2007} and present fast, easy--to--analyze, quantitative measurements to characterize semiconductor spin qubits.  Contrary to the relatively slow tuning based on direct current (DC) electron transport through the dot,\cite{foletti_universal_2009} in our scheme, all scans necessary for characterising a device  can be performed within a few seconds by using pulsed gate measurements and charge sensing with RF readout. Furthermore, as the tuning parameters of interest are obtained directly as fit parameters and require no human intervention, these analysis methods are well suited as a basis for the full automation of the complete tuning procedure.
All measurements presented here were performed on two-electron spin qubits in GaAs quantum dots, but the basic tuning principles can easily be adapted to other quantum-dot-like qubit systems.

The outline of this paper is as follows: In Sec.\,\ref{ch:layout} we  introduce the device layout of the two-electron spin qubit in GaAs, and explain the basics of the experimental setup including the RF-reflectometry circuit. Sec.\,\ref{sec:coarsetuning} covers the first step of tuning the device to either the (2,0)-(1,1) or the (1,1)-(0,2)-charge-transition (the numbers indicate the occupancy in each dot).  Additionally, we describe the tuning of the adjacent quantum dot used for charge sensing of the qubit dots. These methods have hardly changed compared to standard quantum transport measurements\cite{foletti_universal_2009} and will need further refinement\cite{baart_computer-automated_2016} for automation. They are included here for completeness.
In Sec.\,\ref{sec:finetuning} we present our methods to quantitatively characterize and fine-tune the qubit. We first motivate the use of virtual gates, a linear combination of several gates that allows changing specific quantum dot parameters individually.
We continue by describing the characterization of the inter-dot tunnel coupling $t_\mathrm{c}$ and the tunnel couplings to the electron reservoirs, which must be tuned to certain values for the proper operation of the qubit. 
Although these three properties already define the qubit, the gate voltages at which various qubit operations can be carried out still have to be determined. Hence, we additionally provide routines for locating fast--reload points used to initialize the qubit in different states, and the ST$_+$ anti-crossing. The latter point allows us to set up a hardware feedback-loop to polarize and stabilize the nuclear spin bath in the GaAs host material.\cite{bluhm_enhancing_2010}
 
All the data shown in this paper were obtained from the qubit of Refs.\,\onlinecite{botzem_quadrupolar_2016,cerfontaine_feedback-tuned_2016-1}. Tuning methods very similar to those described here were also employed in Refs.\,\onlinecite{shulman_demonstration_2012, dial_charge_2013, shulman_suppressing_2014, nichol_quenching_2015}.

\section{Device layout and experimental setup}
\label{ch:layout}
\begin{figure}[!ht]
	\centering
	\includegraphics[trim = 0cm 0cm 0cm 0cm,width=.5\textwidth]{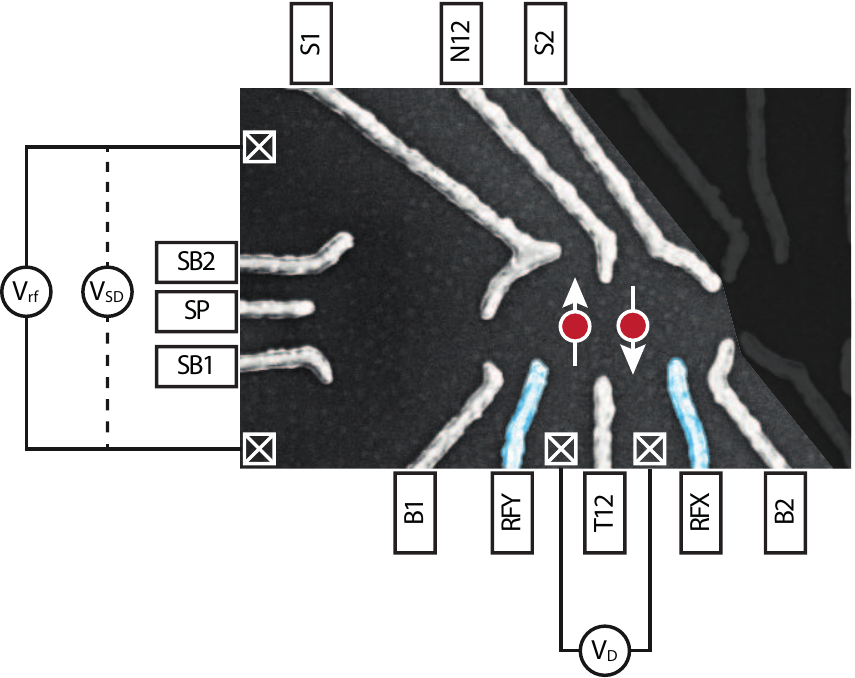}
	\caption[Device Layout.]{{(color online) Device Layout.} The figure depicts the device and the gates used to form the two-electron spin qubit in this work. Applying static voltages to the grey gates confines two electrons in a double dot potential in the 2DEG of a GaAs/Al$_{0.31}$Ga$_{0.69}$As heterostructure. The blue gates, RFX and RFY, are used exclusively for fast manipulation. The dot on the left is used for charge sensing of the double dot, and is embedded in an impedance-matching circuit as the resistive element. The crossed boxes represent ohmic contacts to the leads. The leads close to the RFX and RFY gates are named lead X and lead Y.
	}
	\label{fig:device}	
\end{figure}

The two-electron singlet-triplet spin qubit is defined by the $m_{z}=0$ subspace S$ = \left(\ket{\uparrow \downarrow} - \ket{\downarrow\uparrow}\right)/\sqrt{2}$ and T$_{0}= \left(\ket{\uparrow \downarrow} + \ket{\downarrow\uparrow}\right)/\sqrt{2}$ of two electron spins,\cite{levy_universal_2002} where $\uparrow$ or $\downarrow$ describes the spin state of the electron in one of the dots. These electrons are confined in a GaAs double quantum dot formed by electrostatic gates on top of a two-dimensional electron gas (2DEG), as shown in Fig.\;\ref{fig:device}.

The legacy quantum dot layout, for example from Refs.\,\onlinecite{petta_coherent_2005, elzerman_single-shot_2004}, uses a single gate for high and low frequency control, combined with a bias tee. Home built, on--chip bias--tees, as used in Refs.\,\onlinecite{bluhm_enhancing_2010, foletti_universal_2009}, have the advantage of avoiding difficult-to-model additional inductances in the DC-arm, but still show pulse distortion that needs to be corrected to achieve optimal fidelity between the intended and the actual pulses.\cite{bluhm_enhancing_2010}
To avoid this complication and the resulting pulse imperfections, we use dedicated all--DC static gates and high--frequency control gates. Static voltages of order 1\;V, provided by a home-built voltage source, are applied to the heavily filtered  static gates (depicted in grey in Fig.\;\ref{fig:device}) and are used to define the quantum dots and to tune them into the single-electron regime. In more detail, the broad side gates (denoted S1 and S2) adjust the number of electrons in quantum dot 1 and 2. The barrier gates B1 and B2 control the tunnel coupling to the leads and the inter-dot coupling is controlled by the nose and tail gates, N12 and T12.
Two additional control--gates (named RFX and RFY, depicted in blue in Fig.\;\ref{fig:device}) are used for qubit manipulation by applying mV-scale signals. They are DC-coupled to an arbitrary waveform generator Tektronix AWG5014C, operated at 1\,GS/s. Using separate static and control gates eliminates the need for bias tees and results in a nearly flat frequency response of the control gates from DC to a few hundred MHz, at the cost of one additional gate electrode. The control gates are attenuated by $33$\;dB at various cryogenic stages to reduce thermal noise from room temperature.

A proximal sensing dot is is capacitively coupled to the double dot. The conductance of the sensing dot depends sensitively on the local
electrostatic landscape, so the charge state of the double dot can be read out with it even in regimes where the direct current through the double dot falls below the noise level. \cite{barthel_fast_2010} 
The spin-state of the double dot can also be probed through the sensing dot by spin-to-charge conversion based on Pauli-spin-blockade.\cite{petta_coherent_2005,johnson_singlet-triplet_2005}  The sensing dot is embedded as a resistive component in an impedance matching circuit, so that the conductance through the dot can be monitored using RF-reflectometry \cite{schoelkopf_radio-frequency_1998, reilly_fast_2007, cassidy_single_2007} at a local oscillator frequency of approximately 230\,MHz and a bandwidth of 20\,MHz. We employ a setup similar to Ref.\;\onlinecite{reilly_fast_2007}, with the addition of a cryogenic circulator at base temperature. The demodulated signal $V_\mathrm{rf}$ is a function of the conductance of the sensing dot, and is recorded using an Alazar ATS9440 digitizer board. We typically use a hardware sample rate of 100\,MS/s, which we downsample on--the--fly at full data rate to 250\,kHz using a multithreaded, high throughput C++-based driver for the Alazar card. This downsampled rate arises from a typical length of 4\,$\mu$s for experiments, which usually comprises a 2.5\,$\mu$s long measurement window during which we power the RF-circuit. 

Effects of 1/f-like noise are eliminated from the data by changing the sweep--pulse parameter after each cycle, and then averaging over many repetitions of the parameter sweep to elude slow drifts in the sensor or gate voltage configuration. For a typical tuning dataset, the sweep comprises 100 parameter values and it is repeated 1536 times for a total measurement time of $1536\times 100 \times 4\,\mu\text{s} \approx 600\,\text{ms}$, and then averaged again over $1-5$ repetitions, if necessary. Note that these acquisition parameters are not yet optimized for speed, and we expect that a speed-up of at least a factor of 10 is possible while still maintaining an adequate accuracy of the extracted parameters.

Finally, for the initial coarse tuning of the double dot (see Sec.~\ref{ch:chrg}), instead of using RF-reflectometry, we directly measure the conductance through the double  dot and through the sensing dot. To do so, we apply a voltage bias of 100\;$\mu$V across the devices. The resulting currents are converted to voltages (named $V_\mathrm{SD}$ and $V_\mathrm{D}$ for the sensing dot and double dot, respectively) using a home-built IV-converter and measured with a Standford Research SR830 lock-in amplifier. 

\section{Coarse tuning of the quantum dots}
\label{sec:coarsetuning}
The next two sections describe how to tune the sensing dot and the actual qubit double dot. They do not present novel methods with respect to Ref.~\onlinecite{baart_computer-automated_2016} but are included for completeness.
\subsection{Tuning of the sensing dot}
\label{ch:sensor}
\begin{figure}[!ht]
	\centering
	\includegraphics[trim = 0cm 0cm 0cm 0cm,width=0.5\textwidth]{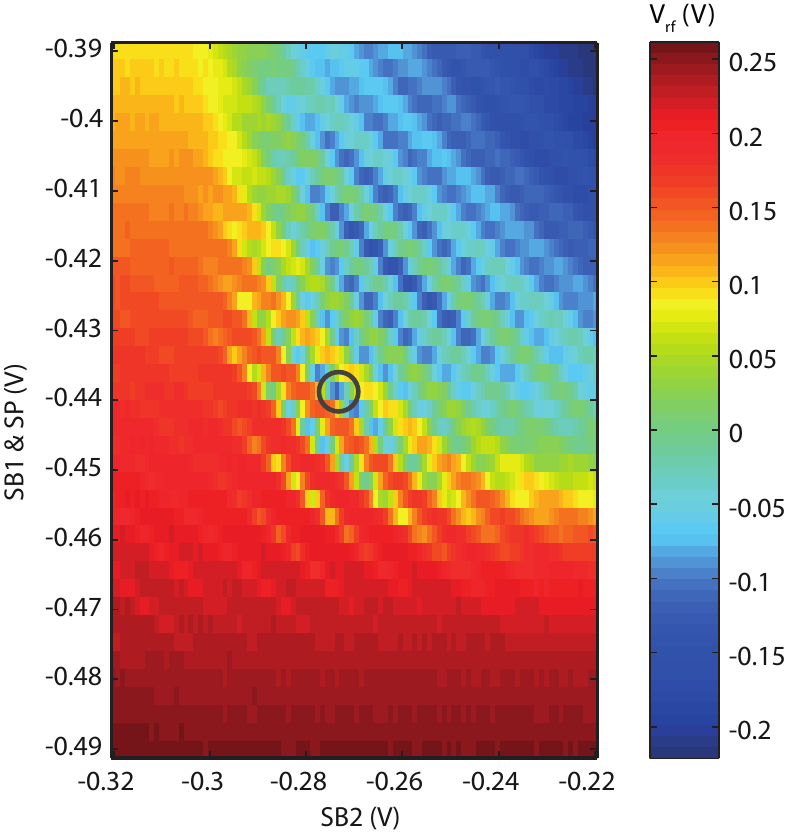}
	\caption[Charge sensor tuning.]{(color online) Charge sensor tuning.  The characteristic charge stability--diagram of the sensing dot, measured with RF-reflectometry. Being proportional to the conductance through the dot, $V_{\rm rf}$, shows Coulomb oscillations once the source and drain barriers are sufficiently opaque. Tuning the sensor to a sensitive position (see circle) allows for charge sensing of the nearby double quantum dot.}
	\label{fig:sensor}	
\end{figure}
The first step in the tuning procedure is to set up charge detection through the sensing dot. This requires that we find a set of voltages applied to the sensing dot gates SB1, SB2 and SP (gate names are defined in Fig.\,\ref{fig:device}) such that the conductance through the dot is maximally sensitive to the local electrostatic potential. To do so, we  measure $V_{\rm rf}$  while performing a two-dimensional scan with the sensing dot gates SB2 vs. SB1\&SP (gate names are defined in Fig.\,\ref{fig:device}). Since $V_{\rm rf}$ is proportional to the conductance through the dot, Coulomb oscillations appear in the measured signal when the applied voltages are sufficiently negative to make the source and drain barriers opaque. Fig.\;\ref{fig:sensor} shows a region in gate voltage space that shows the typical pattern of a single quantum dot.\cite{hanson_spins_2007} In this particular sample, SP and SB1 were shorted and thus had to be kept on the same potential. Usually, SP can be used to fine--tune the dot and to shift it closer to the double quantum dot. To obtain the best charge sensitivity,  the voltages applied to SB2 vs. SB1\&SP have to be tuned to values where the slope of the Coulomb peak is steepest. 

\subsection{Locating the (2,0)-(1,1) or (0,2)-(1,1) charge transition}
\label{ch:chrg}
The second step is to determine the depletion and pinch-off voltages of the different gates that define the qubit double quantum dot. To do that, we directly measure the conductance through the double dot by applying a 100 $\mu$V bias voltage to the leads marked by a crossed box in Fig.~\ref{fig:device}, and measuring the resulting current (converted to the voltage $V_{\rm D}$ by a home-built IV-converter). Measuring the conductance as a function of the voltage applied pairwise to the gates N12 and T12, S1 and B1, S2 and B2 (see Fig.\;\ref{fig:device} for gate nomenclature), allows us to determine the depletion voltages.
\begin{figure}[!ht]
	\centering
	\includegraphics[trim = 0cm 0cm 0cm 0cm,width=0.5\textwidth]{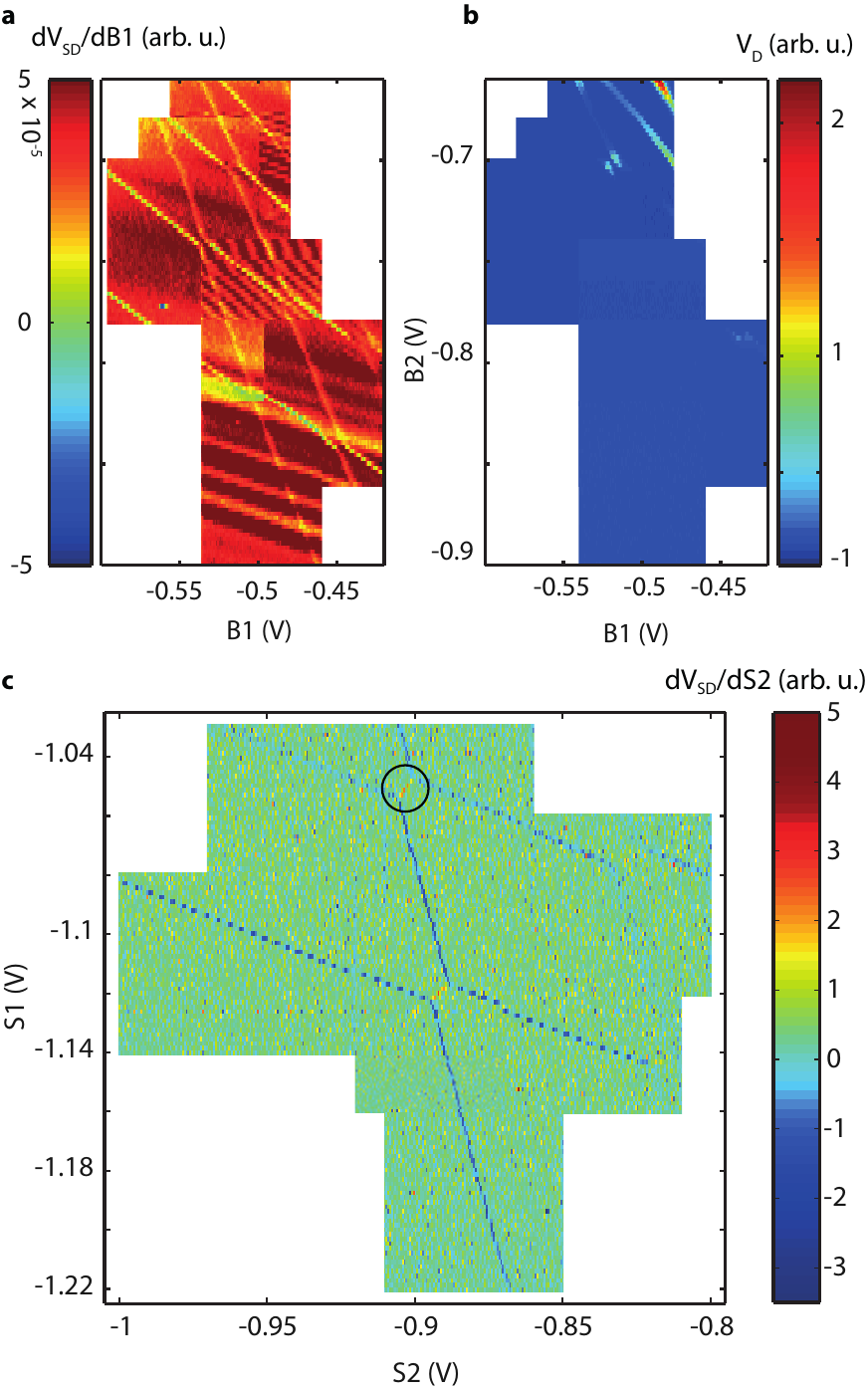}
	\caption[Charge sensing and transport through a double dot.]{(color online) Charge sensing and transport through a double dot. (a) Monitoring the differential conductance through the sensing dot shows the electron occupancy of the double dot in a typical honeycomb pattern. Background oscillations are caused by an imperfect compensation of the sensing dot gates relative to B1 and B2. (b) When measuring transport directly through the double dot, conduction only occurs at the triple-points. Since B1 and B2 also influence the coupling to source and drain, Coulomb peaks become faintly visible only for not too negative voltages. 
(c) Same as in (a), but this time using the side gates S1 and S2 instead of B1 and B2. This typically reduces the background oscillation in the transconductance of the sensing dot.  The intensity of the lines designing the honey-comb pattern reflects the transparency of the tunnel barriers to the external leads, which is influenced  by the side gates S1 and S2	
	}
	\label{fig:chrgst}	
\end{figure}
The gate voltages are then set close to their depletion voltages and the device is tuned close to complete pinch-off. Next, we perform a two-dimensional scan over a couple of tens of mV with the gates B1 and B2. 
Usually we anticipate to first form a large single quantum dot and then separate it into two dots by applying more negative voltages on the T12 and N12 gates. If the tunnel barriers between the dot and leads X and Y are almost pinched off and have similar transmission probabilities, Coulomb blockade peaks should appear, showing the characteristic honeycomb pattern of a lateral double quantum dot.\cite{hanson_spins_2007} 

\begin{figure}[!ht]
	\centering
	\includegraphics[trim = 0cm 0cm 0cm 0cm,width=.5\textwidth]{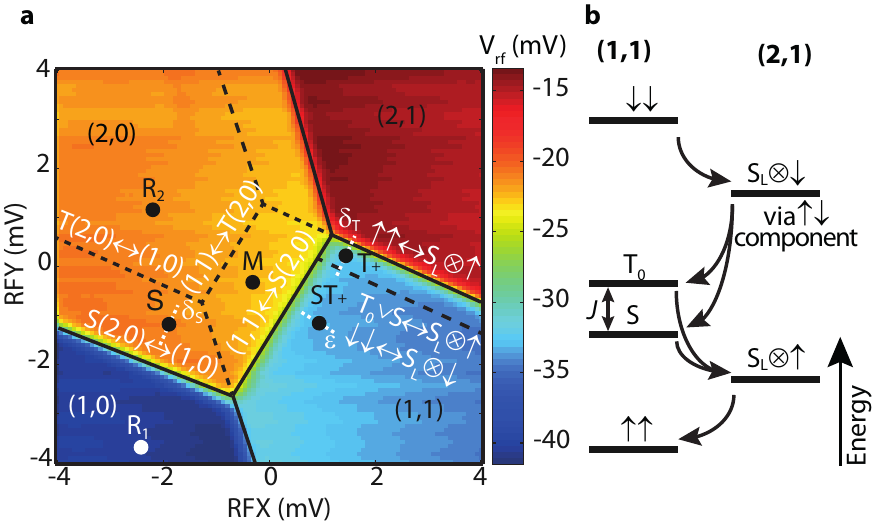}
	\caption[Charge stability diagram.]{(color online) Charge stability diagram. ({a}) High--resolution charge stability diagram around the (2,0)-(1,1)-transition used to define the two-electron spin qubit. Important points that are used to initialize the qubit in different states and for measurement, are marked by dots, and further explained in the main text. Transitions are labeled in white. This diagram has been measured by RF-reflectometry and we subtract a background slope to eliminate direct sensor response. Different charge states in the double--dot  correspond to clearly distinguishable values of $V_{\rm rf}$.   
	({b}) This diagram shows the energy relaxation cascade used to initialize the (1,1) ground state $\ket{\uparrow\uparrow}$ at point T$_+$ (see Sec.~\ref{ch:tpload}). S$_\mathrm{L}\otimes\downarrow(\uparrow)$ denotes the state of a singlet state in the left quantum dot and an down (up)-state in the right dot. The arrows indicate relaxation via electron exchange with the (2,1) charge configuration.}
	\label{fig:chrg3}	
\end{figure}

Observing this honeycomb pattern can be challenging because even though applying negative voltages on the gates B1 and B2 primarily empties the dots, it will eventually also close the tunnel barriers to the reservoirs X and Y, resulting in the current going to zero and hardly detectable Coulomb peaks unless all other gates are carefully tuned, see Fig.\;\ref{fig:chrgst}(b).  To study the double quantum dot with closed leads, we use the sensing dot. Due to the capacitive coupling between the double dot and the sensing dot, a change in the occupation of the double dot results in an abrupt change in the current through the sensing dot and therefore into a sharp signature in the transconductance $d{\rm V_{SD}}/d{\rm B}_{1}$, Fig.\;\ref{fig:chrgst}(a). 
When performing this type of scan, the voltage SB2 is adjusted to compensate the unintentional influence of the stepping gate B2 on the potential of the sensing dot.  Similar scans can also be performed by using the side gates S1 and S2 instead of B1 and B2. This typically reduces the background oscillation in the transconductance of the sensing dot (see Fig.\;\ref{fig:chrgst}c), as the gates S1 and S2 have a weaker influence on the sensing dot than B1 and B2.   
Going towards more negative voltages eventually locates either the (2,0)-(1,1) or the (0,2)-(1,1) charge transition.
Once a suitable transition has been found, we adjust S1 and S2 such that a recorded high-resolution charge stability diagram via RF-reflectometry using the RF-gates, RFX and RFY,  is centered around the transition of interest (see Fig.\;\ref{fig:chrg3} (a)). The fine--tuning of the qubit can now begin.
\section{Fine tuning of the qubit}\label{sec:finetuning}

This section describes the fine-tuning of a two-electron qubit in detail. Sec.~\ref{sec:triplepoints} describes how we automatically locate the triple points in the RFX-RFY plane.
In Sec.\,\ref{ch:virt}, we motivate the use of virtual gates, a linear combination of several gates that allows us to change specific quantum dot parameters individually. 
In Sec.\,\ref{sec:leads} and \ref{ch:tc}, we present a qualitative characterization of the tunnel couplings to the two adjacent electron reservoirs and of the inter-dot tunnel coupling  between the two quantum dots. Sec.\,\ref{ch:measp}-\ref{ch:stp} describes how to locate the fast reload points for S- and T$_+$-states and the location of the ST$_+$ anti-crossing. All measurements presented in this section are performed using RF-reflectometry on the sensing dot.

\subsection{Locating the triple points}
\label{sec:triplepoints}
All the fine--tuning routines discussed in this paper (as well as the operation of a qubit in a actual experiment) require the accurate characterization of  the charge stability diagram. In other words, it is necessary to know the exact position in the RFX--RFY plane of the triple points, the points in the charge stability diagram where three charge states are energetically degenerate (e.g. (1,0),(1,1) and (2,0)), and of the so--called lead transitions, the transitions between different charge--states of the double dot that involve electron exchange with one of the reservoirs (e.g. the transition $(1,0)\leftrightarrow (1,1)$). Previously, extracting this information has been done manually. Here we present instead an automated routine, based on a measurement of the charge stability diagram near the (2,0)-(1,1) transition, followed by fitting to a simple model that allows the extraction of the relevant parameters, namely the location of the two triple points and the slopes of the lead transitions in the RFX--RFY plane. 

\begin{figure}[ht]
	\centering
 	\includegraphics[trim = 0cm 0cm 0cm 0cm]{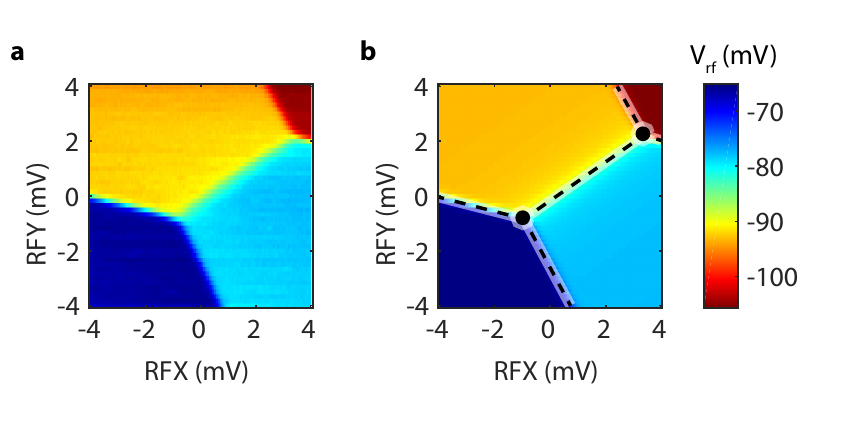}
	\caption[Charge stability diagram fitting.]{{(color online) Fitting of the charge stability diagram.} ({a}) Charge stability diagram measured using RF-reflectometry. Different values of $V_{\rm rf}$ correspond to different charge ground states. ({b}) Fit of the stability diagram using the model described in the main text. Circles and lines mark the automatically detected triple points and the positions of the charge transition. In both figures we subtracted a background value due to the direct influence of the charge sensor extracted from the fit.}
	\label{fig:chargefit}
\end{figure}

 The fitting model consists of two parts. The first part is a two-site Hubbard Hamiltonian without spin:
  \begin{align*}
  \label{eq:tuning:fithamiltonianapprox}
      H =
      \setlength{\arraycolsep}{0pt}
      \begin{pmatrix}
          E_{1,0} + v_1 & 0 & 0 & 0 \\
          0 & E_{1,1} + v_1 + v_2 & t_\mathrm{c} & 0 \\
          0 & t_\mathrm{c} & E_{2,0} + 2 v_1 & 0 \\
          0 & 0 & 0 & E_{2,1} + 2 v_1 + v_2 \\
      \end{pmatrix}.
  \end{align*}
in the charge basis $j \in \left\{(1,0), (1,1), (2,0), (2,1)\right\}$. Here, $E_j$ are the basis state energies at zero voltage, $t_\mathrm{c}$ is the inter-dot tunnel coupling and $v_i$ the on-site potential, which can be calculated knowing the voltages applied to the RF-gates $V_i$ and their respective lever arms, including cross-capacitance. The index $i$ indicates RFX and RFY, respectively ($V_1=$RFX and $V_2=$RFY ). 
The spectrum of this Hamiltonian can be calculated analytically to find the ground state charge configuration at each point in the RFX-RFY plane. Since measurements like those presented in Fig.~\ref{fig:chrg3} and Fig.~\ref{fig:chargefit} are slow compared to the system dynamics, we assume that 
the occupation probability of each state corresponds to thermal equilibrium.  
In this way, for each point in the RFX-RFY plane we can calculate a vector $\mathbf{p}$,  describing the occupation probabilities of the various charge basis states. 

The second part of the fitting model is a linear model for the charge sensor: 
\begin{equation}
	S = \mathbf{p} \cdot \mathbf{s} + s_{\mathrm{ct}, 1} V_1 + s_{\mathrm{ct}, 2} V_2 + S_0 \mathrm{,}
\end{equation}
where $S$ is the charge sensor output, $\mathbf{p}$ is the ground--state charge population vector determined above,  $\mathbf{s}$ is a vector that contains the sensor output for each charge eigenstate, $s_{\mathrm{ct}, i}$ account for direct crosstalk betwen the RF-gates and the sensor, and $S_0$ is an offset.
The components of  $\mathbf{s}$, as well as  $s_{\mathrm{ct}, i}$ and $S_0$,  the lever arms, the cross-capacitances, the energies $E_i$, and the inter-dot tunnel-coupling $t_c$ are treated as fitting parameters, while the input parameter for the fit are the 2D sensor output data and the voltages $V_i$ applied to the RF--gates.  A typical measurement and a fit to the data is presented in Fig.\;\ref{fig:chargefit}. From the fit parameters the position of the triple points as well as the slopes of the leads transitions in RFX-RFY plane are extracted. These values are used as reference points in all the following tuning procedures, and to recalibrate the set-up after a charge rearrangement. 

\subsection{Setting up virtual gates}
\label{ch:virt}

Once the (2,0)-(1,1) (or the (0,2)-(1,1)) charge transition has been located and the voltages applied to the  DC-gates have been adjusted to center the charge stability diagram for scans using the RF-gates, it is advisable to switch to virtual gates. These virtual gates are given by a linear combination of three physical gates (see Tab.\,\ref{tab:virt}) that allow tuning the parameters of the double dot while leaving the charge stability diagram in the RFX and RFY plane unaffected.   
Virtual gates are chosen such that each of them  affects primarily one specific dot parameter:  gates LeadY and LeadX  change the tunnel coupling to the respective lead, while the tunnel coupling between the dots is manipulated by the virtual gates T and N.  In each case, in addition to changing the physical gate that mostly influences the desired parameter, a compensating voltage is applied to the S1 and S2 gates to cancel out any cross-capacitance effect. Virtual gates X and Y depend only on S1 and S2 and are used to readjust features in the charge stability diagram in case of imperfect compensation from the virtual gates or of charge rearrangements. 

In order to obtain the virtual gate coefficients shown in Tab.\,\ref{tab:virt}, we focus on the lead transitions and measure how the potential applied on a certain gate shifts their position in the RFX--RFY plane. To do so, we apply two different voltages (typically differing by 2-6\;mV) to each of the DC gates, in turn. For each set of voltages, we measure the dot's response while sweeping RFX or RFY across both lead transitions, as shown in Fig.\;\ref{fig:tune1}(a) for case of the Y-lead. In these curves, the two plateaux correspond to two different charge states of the double dot. The labels close to each curve indicate the gate on which the potential is changed.  To extract the influence of that gate on the position of the lead transition, the value of RFX (or RFY) at which the transition between the two plateaux occurs, we use a  phenomenological fit-model corresponding to a Fermi distribution,

\begin{align}\label{eq:tuning:fitresp}
	V_\mathrm{rf}(v) = V_\mathrm{rf, 0} +\delta V_\mathrm{rf} v - \frac 1 2 A \left(1+\tanh\left(\frac{v - v_\mathrm{lead}}{w}\right)\right).
\end{align}
Here, $v$ is the voltage on either the RFX or RFY sweeping gate. The first term $V_\mathrm{rf, 0}$ in Eq.\,\eqref{eq:tuning:fitresp} represents the background value of the charge sensing signal $V_\mathrm{rf}$. The linear term $\delta V_\mathrm{rf}v$ accounts for the direct influence of the sweeping gate on the conductance through the sensor. The third term accounts for the excess charge once an electron tunnels into or out of the quantum dot and includes a finite electron temperature via $w$, while $v_{\rm lead}$ defines the position of the lead transition. 
We use $V_\mathrm{rf, 0}, \delta V_\mathrm{rf}, A, w $ and $ v_\mathrm{lead}$ as fit parameters. 
The values of $v_\mathrm{lead}$  extracted from these fits depend on the voltage applied to all the DC-gates, and are used to construct a $2 \times 6$ cross-capacitance matrix. Virtual--gate coefficients are then extracted by inverting the appropriate sub-matrices of the cross-capacitance matrix.  Typical values are given in Tab.\,\ref{tab:virt}. 
The virtual gate coefficients can be further fine-tuned by applying the same principle to study the influence of the DC--gates on the location of the triple points of the (2,0)-(1,1) charge transition or on the position of the ST$_+$ anti-crossing.

A similar concept is used in Ref.\,\onlinecite{baart_single-spin_2016} to perform orthogonal charge stability diagrams in a three-electron quantum dot and in Refs.\,\onlinecite{shulman_demonstration_2012, nowack_single-shot_2011}.

\begin{table}
	\caption{{Virtual gates.}  Typical values for the coefficients of the virtual gates.
	The values correspond to the ratio of change in physical gate voltages to that of the virtual gate.\\}
	\begin{tabular}{cc|c|c|c|c|c|c}
		&\multicolumn{6}{c}{\textbf{Virtual gate}} \\
		&&	LeadY	& LeadX	& T		& N		& X 	& Y   \\\cline{2-8}\cline{2-8}
		\multirow{6}{*}{\rotatebox[origin=c]{90}{\parbox[c]{1cm}{\centering \textbf{Physical gate}}}}&B1		&1  		& 0  	& 0 	& 0		& 0 	& 0 \\\cline{2-8} 
		&S1		&-0.76		& 0.5	& -0.52	& -0.5	& 1.5	& -2.1 \\\cline{2-8} 
		&T12		& 0 		& 0 	& 1		& 0		& 0		& 0 \\\cline{2-8} 
		&N12		& 0 		& 0 	& 0		& 1  	& 0		& 0 \\\cline{2-8} 
		&B2		& 0 		& 1 	& 0		& 0		& 0  	& 0 \\\cline{2-8} 
		&S2		& 0.26 		& -1.1 	& -1.25 &  -0.5	& -3.68	& 1.03  \\ 
		
		\label{tab:virt}
	\end{tabular}
\end{table}
\subsection{Tunnel coupling to the leads}
\label{sec:leads}
The next step in setting up the qubit is the tuning of the tunnel coupling to leads X and Y (Ohmic contacts next to the RF-gates, see Fig.~\ref{fig:device}), which act as electron reservoirs. The coupling to these leads is controlled by the virtual gates LeadY and LeadX, and it must be weak enough to prevent excess $T_1$ relaxation due to cotunneling or thermal activation, while being strong enough to allow fast qubit initialization, within tens of nanoseconds, at the same time. 

\begin{figure}[!ht]
	\centering
	\includegraphics[trim = 0cm 0cm 0cm 0cm, width=0.5\textwidth]{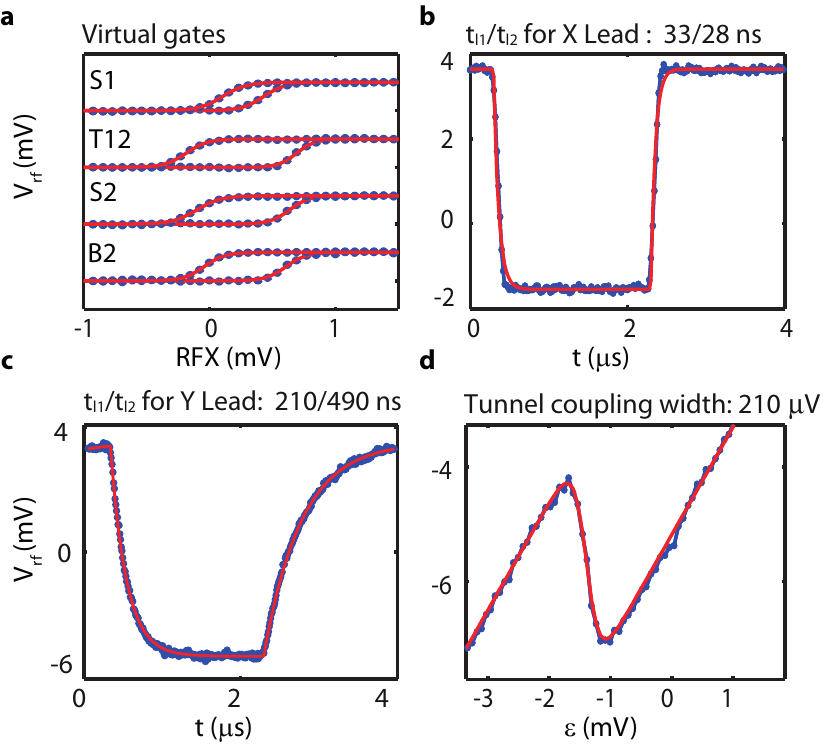}
	\caption[Tuning parameters I.]{ {(color online) Setting up virtual gates and measuring tunnelling times.} 
		({a}) To determine the influence of the various DC--gates on the position of a lead transition, we apply two different voltages to each one of the DC-gates in turn. For each set of voltages, we measure the occupation of the double dot as we sweep across two regions in the charge stability diagram.
	  ({b,c}) To measure the tunnelling times to the leads, we applying MHz voltage--pulses across the respective lead transition, forcing an electron to be exchanged with the respective reservoir. The tunnelling time can be extracted from the rise time of the response signal. ({d}) The inter-dot tunnelling rate can be extracted by sweeping along the detuning $\epsilon$, recording the average charge occupancy and measuring the broadening of the transition.}
	\label{fig:tune1}	
\end{figure}

To extract the tunneling time to the X lead, we apply 25\;MHz square--wave pulses that force the system to switch between the charge states $(1,0)\leftrightarrow (1,1)$ (or, equivalently, between $(2,0)\leftrightarrow (2,1)$), and use the sensing dot to measure the time--dependent occupation of the double dot.
 For this purpose, we average the signal over approximately 1500 periods, recorded at a hardware sampling rate of 100\,MS/s. A typical time trace is shown in Fig.\;\ref{fig:tune1}(b).  Applying the square--wave pulses to regions of charge stability (i.e. where no charge transition is possible) allows us to subtract the background due to direct sensor coupling. The tunnelling time to the lead can be extracted from the rise times of the response to the square pulses, Fig.\;\ref{fig:tune1}(b), with a lower sensitivity bound of about  25\;ns determined by the bandwidth of the tank circuit attached to the sensing dot  (faster tunneling times can be resolved with the reload sweep discussed in Sec.\,\ref{ch:measp}). To fit these data, we use the following model 
\begin{equation}
	V_\mathrm{rf}(t,t_0) =
	\begin{cases}
		V_\mathrm{rf, 0} + \frac 1 2 A \frac{\cosh\left(\frac{t_0}{2t_\mathrm{l,1}}\right)-\exp{\left(\frac{t_0 - 2t}{2t_\mathrm{l,1}}\right)}}{\sinh{\left(\frac{t_0}{2t_\mathrm{l,1}}\right)}} & \text{for } t < t_0 \\
		V_\mathrm{rf, 0} - \frac 1 2 A \frac{\cosh\left(\frac{t_0}{2t_\mathrm{l,2}}\right)-\exp{\left(\frac{t_0 -2t}{2t_\mathrm{l,2}}\right)}}{\sinh{\left(\frac{t_0}{2t_\mathrm{l,2}}\right)}} & \text{for }  t \geq t_0,
	\end{cases}
\end{equation}
where $t_0 = 2\,\mu$s is the half-period of the square pulse. The prefactors and offsets of the exponential rise and decay are chosen such that the curve is continuous.
The same procedure is used to extract the tunnelling times to lead Y, with the only difference that now the square pulse has to force transitions between the charge configurations $(1,0)\leftrightarrow (2,0)$ (or $(1,1)\leftrightarrow (2,1)$). 

Typical target values for $t_{\mathrm{l},1(2)}$ range from 25\;ns to 50\;ns.
Importantly, since all initialization methods addressed here require only tunneling to one lead (see Sec.\,\ref{ch:measp}--\ref{ch:tpload}), the barrier to the other lead can be made less transparent to reduce relaxation. 
\subsection{Inter-dot tunnel coupling}
\label{ch:tc}
The tunnel coupling $t_\mathrm{c}$ between the two dots is mostly controlled by the N and T virtual gates, and determines the strength of the exchange interaction between the two electrons, and therefore the energy splitting $J(\epsilon)$  between the singlet S and the triplet T$_{0}$ in the (1,1) configuration. In order to characterize the tunnel coupling, we measure the broadening of the inter-dot transition between the (2,0)-(1,1) charge configuration by sweeping the detuning $\epsilon$ orthogonally across the (2,0)-(1,1)-transition (see Fig.~\ref{fig:chrg3}(a)), and recording the average charge state \cite{dicarlo_differential_2004} as shown in Fig.\;\ref{fig:tune1}(b). Using pulses from the AWG, we typically measure each detuning step for 1\,$\mu$s and average over 4000 scans for a total measurement time of 0.4\,s. For simplicity, we extract the broadening of the transition by fitting Eq.\,\eqref{eq:tuning:fitresp} to the data, rather than using the physically correct model of an avoided crossing, as we find the difference between the two approaches to be marginal.  {The value extracted for the effective temperature $w$ now represents the inter-dot tunnel coupling $t_{c}$.} Good values for $t_\mathrm{c}$ for the operation of the qubit range from 18\;$\mu$eV to 24\;$\mu$eV, using an estimated lever arm of 9.8\,V/eV. Smaller values of the tunnel coupling would lead to Zener--tunnelling when adiabatically sweeping through the (2,0)-(1,1) transition, and should be therefore avoided. 

This characterization method  is limited by temperature broadening, which, in our set-up 
prevents tunnel couplings below 9\;$\mu$eV to be resolved.  An alternative approach for determining the inter-dot tunnel coupling based on time-resolved charge sensing is described in Ref.\,\onlinecite{gorman_extracting_2016}. Furthermore, the tunnel coupling can also by extracted by photon assisted tunneling spectroscopy.\cite{schreiber_coupling_2011} Compared to the presented method, both alternatives are time-consuming and thus less attractive for our purposes.

\subsection{Locating the measurement point}
\label{ch:measp}
The operation of a qubit relies on the ability to reliably initialize the qubit in a well known state and to accurately measure the qubit's final state.\cite{divincenzo_physical_2000} 
Historically, the standard approach for initializing a spin--qubit in a singlet state is based on the transition cycle T(1,1)$\rightarrow$(1,0)$\rightarrow$S(2,0), which requires electron exchange with both reservoirs.\cite{johnson_singlet-triplet_2005}  Here, we present a modified version that only relies on tunnel coupling to one lead. Compared with the old approach, this procedure requires less tuning and enables simpler future device layouts. 
It also allows for an enhanced charge--detection readout--scheme\cite{studenikin_enhanced_2012} to counteract the visibility loss at high magnetic field gradients.\cite{barthel_relaxation_2012}

\begin{figure}[!ht]
	\centering
	\includegraphics[trim = 0cm 0cm 0cm 0cm,width=0.5\textwidth]{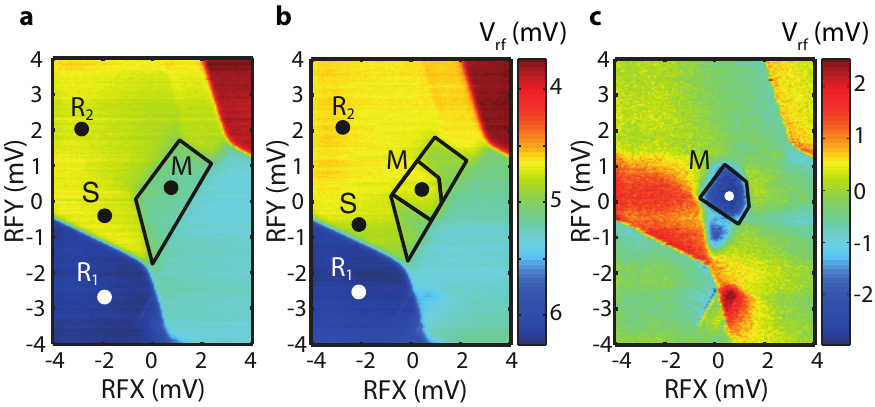}
	\caption[Measurement triangle.]{(color online) Measurement trapezoid. ({a}) To determine the measurement trapezoid, i.e. the region of the RFX-RFY plane where Pauli spin blockade allows for spin--charge conversion, we record a regular charge stability diagram while simultaneously reloading a random (2,0)-spin configuration using the pulse sequence M-R$_{1}$-R$_{2}$-M (see main text). The measurement trapezoid appears as an area where the read-out signal $V_{\rm rf}$ is in between the value typical of the S(2,0)  and of the (1,1) configurations (turquoise--yellowish area). For this specific sample, the blurred boundaries of the readout trapezoid reflect a failure of the random load pulse sequence rather than instabilities. ({b}) Adding a wait time at point S after the pulse sequence from ({a}) maps out the ``mouse-bite'' (yellow triangle within the measurement trapezoid), i.e. the region of singlet reload within the measurement triangle (see main text). ({c}) The difference between {a} and {b} highlights the anticipated readout area for the qubit (blue area).}
	\label{fig:zoom}
\end{figure}

We first need to locate the region of metastable (1,1) triplets within the (2,0) ground state charge configuration (highlighted area around point M in Fig.\;\ref{fig:chrg3}(a)).
To do so, we perform a regular charge--scan,~\cite{barthel_rapid_2009,johnson_singlet-triplet_2005} i.e  we repeatedly apply the pulse scheme M-R$_1$-R$_2$-M, while sweeping through the RFX-RFY plane by adding a DC-offset to the RF-gates. At points R$_{1}$ and R$_{2}$ we wait for $200$\;ns.
Data acquisition is masked--out during the pulse sequence, and we read out the state of the system only at the final point M. If, during a scan, point M falls deep into one of the charge--stability regions, we then  simply observe the same response as in Fig.~\ref{fig:chrg3} and~\ref{fig:chargefit}. However, if the pulse sequence  M-R$_1$-R$_2$-M drives the system through three stability regions as indicated in Fig.~\ref{fig:zoom}(a), the measured signal will then have a value between the one characteristic of the (2,0) charge state and the one characteristic of the (1,1) state. The reason is that when we step from R$_{1}$ to R$_{2}$, we initialize at random either a singlet S(2,0) or a triplet T(2,0) state. If the system is in the T(2,0) state, then it tunnels into the (1,1) configuration when we step back to point M. In comparison, if the system in R$_{2}$ is in the S(2,0) state, it remains in this state. 
In this way we map out the so-called measurement triangle (or trapezoid, if the singlet-triplet splitting is smaller than the inter-dot charge coupling, as in  Fig.~\ref{fig:zoom}).

To determine the position of the singlet reload point, we extend the pulse scheme to  M-R$_1$-R$_2$-M-S-M (see Fig.~\ref{fig:zoom}(b)), including an additional 100\;ns pause at point S. When point S stays energetically between the (1,1)-T(2,0) and (1,1)-S(2,0) transitions (see Fig.~\ref{fig:chrg3}), then electron exchange with the Y-reservoir will lead to the initialization of a (2,0) singlet state.  If this is the case, measuring the state of the system back at point M will give a value of $V_{\rm rf}$ typical of the (2,0) charge state, instead of the intermediate value observed with the M-R$_1$-R$_2$-M pulse scheme. Scanning the position of the pulse cycle over the RFX-RXY plane (the relative position of the points M, R$_{1,2}$ and S is kept fixed) maps out the area known as ``mouse bite'', visible in see Fig.~\ref{fig:zoom}(b). This area becomes even more clearly visible if one subtracts the results of the two scans, see Fig.~\ref{fig:zoom}(c). Here, the blue area indicates a region of the charge stability--diagram where both measurements based on Pauli spin blockade and singlet initialization are possible.  

Once the ``mouse bite'' has been identified, we also know a suitable position of point S for fast initialization of the qubit in the S(2,0) state. To further optimize the position of S, we repeat the pulse sequence M-R$_1$-R$_2$-M-S-M, but now sweeping the position of point S perpendicularly to the (2,0)-(1,0) transition--line, while keeping all other points of the sequence fixed.  In particular, point M has to lay within the ``mouse bite''. As before, we measure the state of the system only in the final point M.  The response signal $V_{\rm rf}$ shows a plateau as a function of the position of point S, at the signal-level of the (2,0) charge ground state, see Fig.~\ref{fig:tune2}(a). The two ridges where the signal increase represent the onset of the transitions (1,0)$\rightarrow$ S(2,0) and (1,0)$\rightarrow$ T(2,0), respectively.  The optimal position of point S for the operation of the qubit lays symmetrically between these two points. 

In a last characterization scan, we fix point S at the optimal position and repeat the pulse sequence  M-R$_1$-R$_2$-M-S-M now varying 
the waiting time $t$  at point S. Again, we measure the state of the system only in the final point M. The longer we wait in point S, the higher the probability to initialize a singlet $S(2,0)$, and therefore the lower the value of V$_{\rm rf}$ measured at point M, see Fig.\ref{fig:tune2}(b).  Fitting these date with a simple exponential decay  
\begin{align}
	V_\mathrm{rf}(t) = V_\mathrm{rf, 0} +A e^{-\frac{t}{t_\mathrm{load}}},
\end{align}
where $V_\mathrm{rf, 0}, A$ and $t_\mathrm{load}$ are fit parameters. For a well--tuned dot, the singlet reload--time $t_\mathrm{load}$ typically lies in the range of 10 to 50\;ns. This characterization scan is complementary to the one presented in Sec.\,\ref{sec:leads}. It exploits the full time resolution of 1\,ns of the AWG, as it is not limited by the bandwidth of the readout tank circuit. 

Having identified an optimal singlet--reload point and characterised the singlet reload--time $t_\mathrm{load}$, in the rest of this paper whenever we will be talking about ``intitalizing the qubit in the singlet S(2,0) state'' we mean the following procedure: i) go to the optimal point S, ii) wait in this position for $\sim 5\, t_{\rm load}$, iii) move to measurement point M. Note that this initialization procedure requires only electron exchange with one lead, which means that only one tunnel barrier has to be tuned to find an optimal operation regime. Moreover, the initialization time is simply given by the tunnel coupling to this lead, and can be as fast as a few tens of nanoseconds.  

\subsection{Locating the triplet T$_+$ reload point}
\label{ch:tpload}
A fundamental technique for the operation of qubits based on GaAs is dynamical nuclear polarization (DNP). This technique is used for stabilizing the surrounding bath of nuclear spin, and relies on the ability to initialize the (1,1) ground state T$_+$.\cite{bluhm_enhancing_2010}
Traditionally, this is done exploiting both the  (2,1)-(1,1) and the (1,0)-(1,1) transitions, i.e. allowing electron exchange with both leads, see Ref.\,\onlinecite{foletti_universal_2009}.  Here we report a different approach, again based on tunneling only to one lead. The trick is to exploit the relaxation cascade shown in Fig.\;\ref{fig:chrg3}(b), which characterizes the region of the stability diagram close to the (1,1)-(2,1) transition. In the presence of an external magnetic field $B_\mathrm{ext}$, the triplet $T_+$ represents the ground state of the (1,1) charge configuration, and transitions from the (2,1) ground-state to the excited states of the (1,1) configuration are not energetically allowed close to the (1,1)-(2,1) boundary. 
This means that if we initialize the qubit in the S(2,0) state, and then pulse to a point T$_+$ close to the (1,1)--(2,1) transition, (see Fig.\;\ref{fig:chrg3}(a))  the qubit either ends up directly into the T$_{+}$ state, or it will eventually reach this state at the end of the relaxation cascade sketched in Fig.\;\ref{fig:chrg3}(b).  Importantly, for this to happen,  we need i) not to cross the upper triple point during the S $\to$ T$_{+}$ pulse, in order not to introduce  measurements artifacts; ii) to ensure that the exchange interaction satisfy the requirement $B_\mathrm{ext}>J(\epsilon)>\Delta B_z$, which is necessary for having sufficient mixing between the $\ket{\uparrow,\downarrow}$, $\ket{\downarrow,\uparrow}$ states and the full relaxation to the T$_+$ ground state. Here $\Delta B_z$ is the difference of magnetic field in the two dots.
\begin{figure}[!ht]
	\centering
	\includegraphics[trim = 0cm 0cm 0cm 0cm, width=0.5\textwidth]{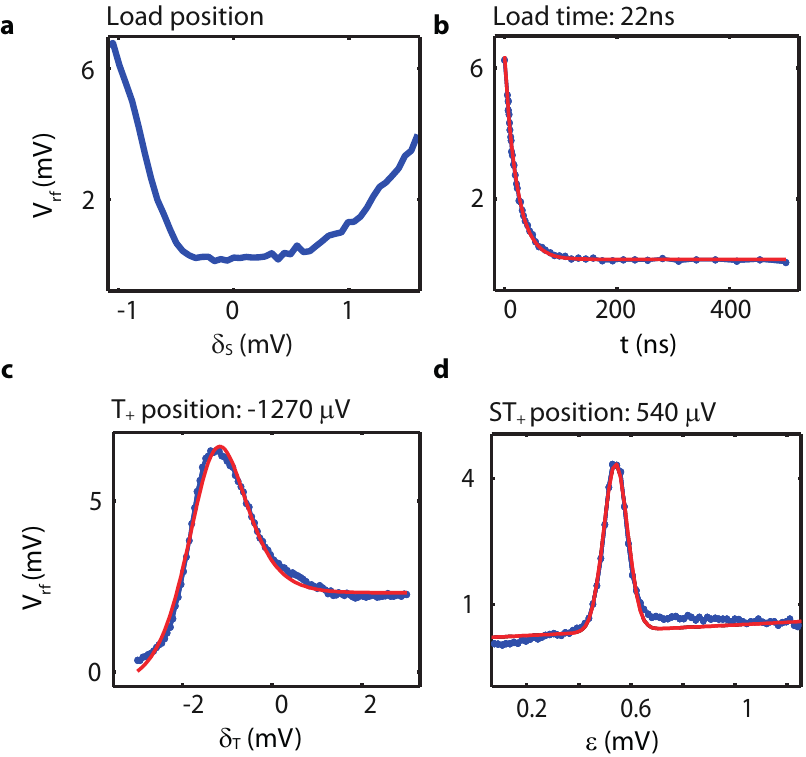}
	\caption[Tuning parameters II.]{{(color online) Locating the operational points of the qubit.} ({a}) To determine the optimal position of the singlet reload point, we shift the position of point S in the pulse sequence M-R$_1$-R$_2$-M-S-M along the direction $\delta_\mathrm{S}$ (see text). The optimal position for S, is in the middle of the plateau of low $V_{\rm rf}$ values. {(b)} The singlet reload--time is extracted by applying the pulse sequence M-R$_1$-R$_2$-M-S-M, and measuring the triplet return--probability as a function of the waiting time $t$ at point S. 
 	{(c,d)} The positions of the T$_{+}$ reload--point and of the ST$_{+}$ transition can be determined using the pulse sequences described  in Sec.~\ref{ch:tpload} and \ref{ch:stp} respectively, and result in peaks in the measured $V_{\rm rf}$ signals as function of the displacement $\delta_{\rm T}$ and of the detuning  $\epsilon$.}
	\label{fig:tune2}	
\end{figure}

In order to find the optimal T$_+$ reload point in the charge stability diagram, we perform the following sweep. We initialize the qubit in the S(2,0) state and then pulse to point T$_+$ without crossing the upper triple point to avoid measurement artefacts. The distance between T$_{+}$ and the upper triple--point has to be chosen such as to fulfil the energy requirement $B_\mathrm{ext}>J(\epsilon)>\Delta B_z$. After a waiting time of 100\,ns to allow energy relaxation, we switch back to the measurement point M and measure the state of the system. We repeat this procedure while sweeping the position of point T$_+$ by $\delta_{\rm T}$, perpendicularly to the direction of the (1,1)$\to$(2,1) transition. The optimal position of point T$_{+}$ appears as a maximum of $V_{\rm rf}$ as function of $\delta_{\rm T}$ (see Fig.\;\ref{fig:tune2}(c)), indicating that indeed a triplet was initialized while waiting at point T$_{+}$.
To extract the exact position of the reload point, we use a phenomenological model motivated by Eq.\,\eqref{eq:tuning:fitresp} and given by
\begin{align}
	V_\mathrm{rf}(\delta_\mathrm{T})= V_\mathrm{rf, 0} + &\frac 1 2 A_1 \left(
	1+\tanh{\left(\frac{\delta_\mathrm{T} - \delta_\mathrm{tl,1}}{w}\right) } \right) \nonumber\\
	- &\frac 1 2 A_2 \left(
	1+\tanh{\left(\frac{\delta_\mathrm{T} - \delta_\mathrm{tl,2}}{w}\right) } \right)
\end{align}
to fit the data. The position of the T$_+$ point is then given by $(\delta_\mathrm{tl,1}+\delta_\mathrm{tl,2})/2$.

\subsection{Locating the ST$_+$-transition}
\label{ch:stp}

In addition to the location of the T$_{+}$-reload point, to perform DNP it is necessary to also know the location of the S-T$_+$-anti-crossing. To find the latter, we follow Ref.\,\onlinecite{petta_coherent_2005} and  initialize the qubit in the singlet state, then change the detuning $\epsilon$, wait 100\;ns at a given detuning, and then return to the measurement point and read out the final qubit state. When the detuning is at the S-T$_+$ anti-crossing, the hyperfine and the spin-orbit interaction can turn the initialized S-state into an T$_{+}$-state, giving rise to a maximum in the measured $V_{\rm rf}$ as a function of $\epsilon$.  Because the location of the ST$_+$ transition strongly depends on the local magnetic field, any unintentional  polarization, for example, due to hyperfine mediated spin flips at the ST$_+$ transition, shifts the precise position of the anti-crossing. To avoid this problem, we include pauses of a few milliseconds at the end of each $\epsilon$-sweep, to allow any unintentional polarization to relax. If needed, we  average over a few different sweeps, and fit our data with a Gaussian model
\begin{align}\label{eq:tuning:stp}
	V_\mathrm{rf}(\epsilon)= V_\mathrm{rf, 0} + \delta V_\mathrm{rf}\epsilon + A e^{-\frac{(\epsilon - \epsilon_\mathrm{stp})^2}{2w^2}} 
\end{align}
to extract the position $\epsilon_\mathrm{stp}$ of the ST$_+$ transition ($V_\mathrm{rf, 0}$, $\delta V_\mathrm{rf}$, $A$, $\epsilon_\mathrm{stp}$ and $w$ are fit parameters).

Not only is this position crucial for the pulsed DNP scheme but, in combination with the T$_+$ reload point, it is also used as an anchor point in the charge stability diagram. Adjusting the dot using the X and Y virtual gates to obtain the same values for the ST$_+$ and T$_+$ scan after a small charge--switching event usually restores all quantum dot parameters, and results in the same $J(\epsilon)$ relation. Furthermore, the position of the ST$_+$ crossing is used to automatically determine switching events\cite{buizert_insitu_2008, kuhlmann_charge_2013, pioro-ladriere_origin_2005} that shift the whole transition by several~mV.
\subsection{Tuning workflow}\label{sec:workflow}
To summarize, once the (2,0)-(1,1) charge transition has been identified, the typical fine-tuning workflow of a ST$_{0}$ qubit starts by defining the virtual gates. The next step is to bring the tunnel couplings to the leads in the right regime. As initialization and readout are parts of any of the following pulse sequence (and of any experiment in general), 
the next step of  the workflow requires tuning the singlet reload point S and the measurement triangle.
The energy splitting $J(\epsilon)$ of the qubit is subsequently tuned by adjusting the inter-dot tunnel coupling. A working scan of the S-T$_{+}$ transition as described in Sec.\,\ref{ch:stp} is a good indicator of a suitable inter-dot tunnel coupling for the operation of the qubit. Usually, tuning the tunnel couplings is an iterative procedure, as adjusting the T and N virtual gates used to tune the inter-dot coupling  also affects the coupling to the leads a little. Finally, the position of the T$_+$ reload point is determined.
During the whole tuning procedure we periodically check the exact position of the (2,0)-(1,1) transition by recording a charge stability diagram. The triple points, which act as anchor points, are extracted automatically by either image recognition or by a fit that includes a model of the charge transition, as described in Sec.\,\ref{sec:triplepoints}. The lower triple point is used as a reference for the  measurement point, and either an offset on the RF-gates or on the virtual gates X and Y is used to center the transition accordingly.
The overall fit stability of all scans requires a meaningful signal-to-noise ratio that we phenomenologically find to be on the order of 5 (measured as the ratio of a transition step size to the rms-fluctuation away from the transition in a charge stability diagram ). To ensure a high sensitivity, we periodically check the sensing dot position (see Sec.\,\ref{ch:sensor}) by performing line scans through the charge stability diagram in Fig.\,\ref{fig:sensor}) and adjust the sensor dot gate voltages accordingly.
Manual retuning to restore the quantum dot parameters once the charge sensor becomes insensitive or a charge rearrangement occurs takes in general a few iterations of performing the various characterization scans and adjusting the gate voltages, and can be typically performed in a couple of minutes.

\section{Conclusion}

This paper provides a detailed description of tuning and characterization routines that we use to realize a ST$_0$ qubit in a GaAs double quantum dot. We describe efficient methods to determine the tunnel couplings between the dots and to the leads, and methods to locate the various points in the charge stability diagram that are  needed for the operation of the qubit itself or for pulsed feedback DNP.

While all relevant quantitative double dot parameters are already obtained automatically, the decision of how to adjust the gate voltages is currently man-made by the operator, based on experience. A logical next step is to also automate this step. One complication is that the effect of the T and N gates on the inter-dot tunnel coupling changes substantially in different regions of gate voltage space, or when charge rearrangements in the vicinity of the dot occur, including even sign changes. This behavior will likely render tuning algorithms based exclusively on pre-calibrated gradient information ineffective, requiring more sophisticated, adaptive approaches. 

Nevertheless, we are optimistic that the procedures described here could be used as a starting point for reaching that goal. Improved sample designs\cite{baart_computer-automated_2016,zajac_scalable_2016} and lower disorder that make the response to gate voltage changes more predictable will greatly simplify the task. Complementary to that, self-calibrating approaches such as the use of a Kalman\cite{kalman_new_1960} filter to track the response tensor over the recent tuning history appear promising. Such advances will be indispensable as soon as the number of qubits increases substantially.

\subsection*{Acknowledgements}
This work was supported by the Alfried Krupp von
Bohlen und Halbach Foundation, DFG grant BL 1197/2-1, BL 1197/4-1, SFB 689, the Excellence Initiative of the German federal and state governments, the Deutsche Telekom Foundation, the United States Department of Defense, the Office of the Director of National Intelligence, Intelligence Advanced Research Projects Activity, and the Army Research Office grant W911NF-11-1-0068. S.P.H. was supported by the Department of Defense through the National Defense Science Engineering Graduate Fellowship Program.
We thank F. Haupt for helpful input on the manuscript. We acknowledge support by the Helmholtz Nano Facility (HNF) at the Forschungszentrum J\"ulich.~\cite{HNF} 



%

\end{document}